\documentclass{article}
\usepackage{arxiv}

\usepackage[utf8]{inputenc}
\usepackage[english]{babel}
 
\usepackage{url}            % simple URL typesetting
\usepackage{natbib}
\usepackage{graphicx}
\usepackage{amsmath}
\usepackage{enumitem}

\title{Characterizing the energetics of \\ multi-scale asymmetries during tropical cyclone \\ rapid intensity changes}

\author{
  S. Bhalachandran$^{1,*,**}$,
\ D. R. Chavas$^{1}$,
\ F. D. Marks Jr.$^{2}$,
\ S. Dubey$^{3}$,
\ A. Shreevastava$^{4}$,
\ T. N. Krishnamurti$^{5,+}$ \\ \\
\ 1. Dept. of Earth,  Atmospheric, and Planetary Sciences, Purdue University, West Lafayette, IN, USA
\\
\ 2. Hurricane Research Division, NOAA/AOML, Miami, FL, USA
\\
\ 3. Indian Institute of Technology, Delhi, India
\\ 
\ 4. Dept. of Civil Engineering, Purdue University, West Lafayette, IN, USA
\\ 
\ 5. Florida State University, Tallahassee, FL, USA
\bigskip \\ 
*Corresponding author: saipb@stanford.edu \\
**Current Affiliation: Department of Earth System Science, Stanford University, CA, USA \\
+Deceased Author
}

\begin{document}
\maketitle

%\linenumbers

\begin{abstract}
Our collective understanding of azimuthally-asymmetric features within the coherent structure of a tropical cyclone (TC) continues to improve with the availability of more detailed observations and high-resolution model outputs. However, a precise understanding of how these asymmetries impact TC intensity changes is lacking. Prior attempts at investigating the asymmetric impacts follow a mean-eddy partitioning that condenses the effect of all the asymmetries into one term and fails to highlight the differences in the role of asymmetries at different scales. In this study, we present a novel energetics-based approach to analyze the asymmetric impacts at multiple length-scales during periods of TC rapid intensity changes. Using model outputs of TCs under low and high shear, we compute the different energy pathways that enhance/suppress the growth of multi-scale asymmetries in the wavenumber (WN) domain. We then compare and contrast the energetics of the mean flow field (WN 0) with that of the persistent, coherent vortex-scale asymmetric structures (WNs 1,2) and the more local, transient, sub-vortex-scale asymmetries (WNs $\geq$ 3). We find in our case-studies that the dominant mechanisms of growth/decay of the asymmetries are the baroclinic conversion from available potential to kinetic energy at individual scales of asymmetries, and the transactions of kinetic energy between the asymmetries of various length-scales; rather than the barotropic mean-eddy transactions as is typically assumed. Our case-study analysis further shows that the growth/decay of asymmetries is largely independent of the mean. Certain aspects of eddy energetics can potentially serve as early-warning indicators of TC rapid intensity changes.
\end{abstract}

% \input{Sections/Intro}
% \input{Sections/Methods}
% \input{Sections/TangMom}
% \input{Sections/ScaleInt}
% \input{Sections/Conclusions}

% Intro 

\noindent
\section{Introduction}
Advancements in airborne radar technologies \citep{marks1984airborne,marks1992dual}, satellite remote sensing \citep{chen2006effects}, and high-resolution 3D modeling (e.g., \citealt{anthes1972development,van2008tropical,gopalakrishnan2011experimental}), have provided ample evidence that strong \textit{azimuthal asymmetries} (hereafter, just asymmetries) are inherent to the flow evolution of a tropical cyclone (TC) vortex. 
These asymmetric features are typically present during all phases of a TC's life cycle but are most predominant every time the vortex undergoes a notable transition in its structure and intensity. The growth and decay of these asymmetries are of great scientific interest since their dynamics are strongly linked to the intensification or weakening of the TC vortex \citep{braun2002cloud,hendricks2004role,montgomery2006vortical,van2008tropical}. However, even with the development of state-of-the-art three dimensional, cloud-resolving models that inherently capture the \textit{effect} of these asymmetries, a precise understanding of how these asymmetries impact TC intensity changes is lacking. This is because asymmetries arise from different sources, occur at different spatial and temporal scales, and their evolution in time is influenced by several mechanisms simultaneously. As a result, there are multiple possible pathways for an asymmetric TC vortex and many configurations that can result in either intensification or weakening.

Asymmetries were previously associated solely with the \textit{weakening} of TCs \citep{nolan2003nonhydrostatic,nolan2007tropical,persing2013asymmetric}. However, recent studies \citep{persing2013asymmetric,smith2017dynamics,leighton2018azimuthal} have demonstrated that asymmetries can have a positive or negative role towards TC intensity changes. A natural question that follows such a conclusion is: What are the scenarios under which asymmetries act to positively influence a change in TC intensity and what are the alternative scenarios under which they aid in the demise of the TC? To address this question, we must first characterize the various spatial, spectral, and temporal aspects of asymmetries in the context of TC intensity changes. 

Asymmetries within a TC vortex may arise via two distinct pathways: (i) external and (ii) internal. Externally-induced asymmetries occur when a vortex responds to a significant change in the environmental flow-field (e.g., wind shear or landfall). A sheared, tilted vortex, for example, experiences a series of asymmetric dynamic-thermodynamic reorganization in convection, moist-entropy, and the flow-field \citep{jones1995evolution,frank2001effects}. Such reorganization manifests as coherent, persistent vortex-scale asymmetries. Alternatively, internal vortex-dynamics may result in localized sub-vortex-scale asymmetries \citep{nolan2007tropical,yang2007effect,marks2008structure}. Examples include vortical hot towers \citep{riehl1961some,hendricks2004role,guimond2010multiscale,gopalakrishnan2011experimental} and rain bands \citep{willoughby1984stationary,didlake2013convective}. These are a result of instabilities on multiple scales, i.e., baroclinic, barotropic, convective, etc. (\citealt{willoughby1984stationary,schubert1999polygonal,kossin2000unstable,kossin2001mesovortices}). Although these localized asymmetries cover only a small portion of the TC inner-core, they contribute to a majority of the upward mass transport \citep{braun2002cloud}.  Furthermore, the gradients associated with the asymmetric distributions within the vortex result in eddy fluxes and wave asymmetries that serve to redistribute various quantities such as vorticity, momentum, and moist-entropy \citep{willoughby1977inertia,montgomery1997theory,chen2001spiral,corbosiero2006structure,hendricks2010spontaneous,moon2010gravity}. Ultimately, the dominance of the internal or externally-induced asymmetries is a function of the strength and nature of the environmental flow field; and the vortex's resilience to an external forcing. 

In reality, externally and internally-induced asymmetries coexist within a TC vortex. As a result, asymmetries are generated and persist across many length-scales. For example, the convective entities within a TC vortex range from individual clouds of length-scales $\leq$ 1 kilometer \citep{krishnamurti2012impacts} as well as localized, `tornado-scale' features \citep{wurman2018role,wu2018prevalence} to coherent, vortex-scale convective entities of the order of hundreds of kilometers \citep{ooyama1982conceptual,krishnamurti2005hurricane,houze2009convective,ryglicki2018unexpecteda}. While the vortex-scale asymmetric structures are essentially the result of individual cloud elements organized with other cloud elements in the azimuth, their behavior can be drastically different from their individual counterparts \citep{krishnamurti2005hurricane,ryglicki2018unexpectedb}.

Temporally, asymmetries of different length-scales are associated with different levels of predictability \citep{judt2016predictabilitya,judt2016predictabilityb,finocchio2017predictability}. The lower wavenumbers (WNs; larger scales) have higher predictability limits as opposed to the higher WNs. Processes at scales that are more stochastic (smaller predictability limit) may or may not be persistent enough to influence a vortex-scale transition. Therefore, the inclusion of all such processes into numerical models may or may not offer any advantage in our ability to forecast TC intensity changes. Given the existing bed of axisymmetric theories (e.g., \citealt{emanuel1986air,chavas2015model}), it behooves us identify the magnitude and nature of the impact of asymmetries at multiple length-scales, and then evaluate if their inclusion justifies the subsequent increase in computational cost. On the other hand, ensemble model studies have shown that the smaller-scale processes significantly impact the \textit{timing} of occurrence of intensity changes \citep{zhang2013effects,judt2016predictabilityb,munsell2017dynamics}. This suggests that smaller-scale processes may indeed play an important role in influencing TC intensity changes. However, minimal work has been done thus far to understand these details.

Prior attempts at investigating the asymmetric impacts have relied on a mean-eddy partitioning. Unfortunately, such approaches condense the effect of all the asymmetries into one term and fail to highlight the differences in the role of asymmetries at different scales \citep{saltzman1957equations}. Furthermore, we find that in certain scenarios, the dynamics and thermodynamic asymmetries may counteract each other, which renders forecasting difficult. As an alternative, in this study, we present a novel energetics-based framework that allows us to analyze the impact of asymmetries at multiple length scales. Our emphasis is on periods of rapid intensity changes that are characterized by TC intensity changes of $\pm$30 knots or greater within 24 hours; comprising rapid intensification (RI), and rapid weakening (RW)  (\citealt{kaplan2003large,kotal2013large,wood2015definition}). Specifically, we address: (i) What are the various energy pathways that enhance or suppress the growth of multi-scale asymmetries within the vortex? (ii) What is the relative importance of these energy pathways to one another? (iii) How are the growth and disruption of the asymmetries consequently linked to the rapid intensity changes in numerical simulations of sheared and low-sheared TCs? In the process of deriving a more fundamental understanding of the behavior of asymmetries within TC vortices, we identify specific aspects of their energetics that can potentially be used as early-warning indicators of rapid intensity changes. 

\section{Methods}
Three TCs that underwent rapid intensity changes during their life cycle are chosen for this study. These TCs are  Phailin and Lehar from the 2013 Bay of Bengal cyclone season, and Hurricane Harvey from the 2017 Atlantic hurricane season. Phailin and Harvey serve as examples of TCs that rapidly intensified in low-sheared environments over the ocean and Lehar serves as an example that rapidly weakened over the ocean in a sheared environment. The choice of TCs in low-sheared and sheared environments allows us to compare the cases where the dominant sources of asymmetries are the internal vortex dynamics with those where the asymmetries are externally induced by shear-vortex interactions.  

Simulations from the Hurricane Weather and Research Forecasting (HWRF) model were used for all the TCs of interest. The model is non-hydrostatically mapped on a rotated latitude-longitudinal, Arakawa E-staggered grid with a storm centered hybrid (sigma-p) coordinate in the vertical direction. HWRF was developed by the National Centers for Environmental Prediction (NCEP) and the Hurricane Research Division (HRD) of the Atlantic Oceanographic and Meteorological Laboratory and is updated continuously \citep{Biswas2016hurricane}. 

The model has three nested domains with a grid spacing of 18 km, 6 km, and 2 km respectively. The 2 km output from the inner-most domain of the HWRF version 3.8 were utilized using the same configuration as \citet{alaka2017performance}. There are 75 vertical levels, with 11 levels below 850 mb for adequate resolution of the TC boundary layer (BL). The model employed a combination of Ferrier-Aliago microphysics, the scale-aware Simplified Arakawa-Schubert (SAS) scheme, the Rapid Radiative Transfer Model-G (RRTMG), the Global Forecast System (GFS) Eddy-Diffusivity Mass Flux BL scheme, Noah land surface model, and Geophysical Fluid Dynamics Laboratory (GFDL) surface physics.

\begin{figure*}[htbp]
 \centerline{\includegraphics[width=\linewidth]{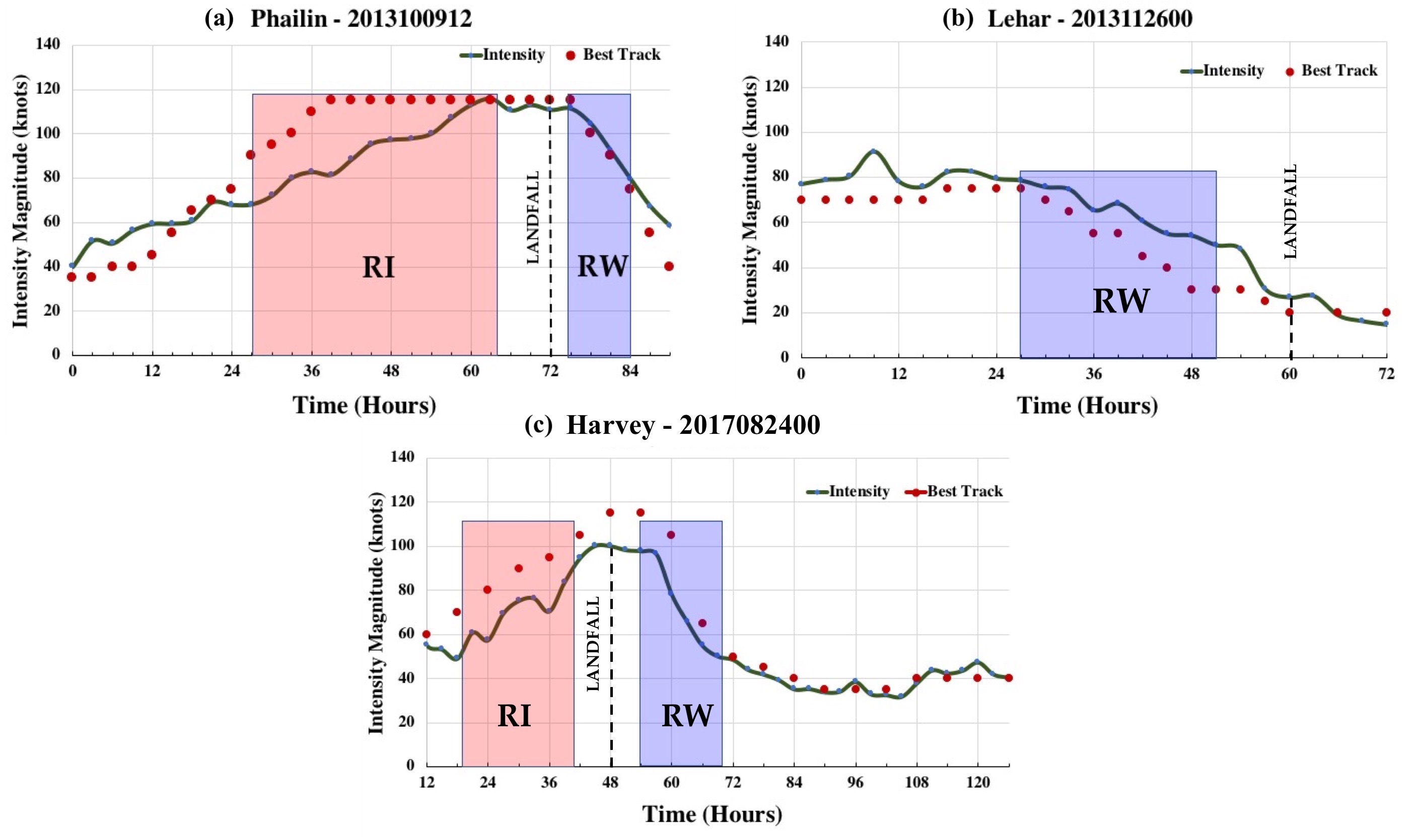}}
  \caption{Time-series plot of intensity for (a) Phailin (initialized on 2013100912)  (b) Lehar (initialized on 2013112600) and Harvey (initialized on 2017082400). The solid line represents the forecast from HWRF and the dots represent the best track intensities. The RI periods for the simulations are shown as red, shaded periods; the RW periods are illustrated as blue, shaded periods and the landfall times are indicated as dashed lines.} 
\label{Figure1_TCintensity.jpg}
\end{figure*}

The inner-domain HWRF output for each of the TCs was then transformed to a storm-centric, cylindrical coordinate system (using a monotonic bi-cubic interpolation) with a grid resolution of $\Delta r = 1 km$ and $\Delta \theta = 1\deg $. We use the location of surface-minimum pressure at each time to calculate the center of the cylindrical coordinate system \footnote{The choice of center can have a significant impact on the generation of asymmetries. For example, in highly sheared environments where the vortex is tilted, a different choice of center may result in a WN-1 asymmetry at a certain range of radii. A change in the center with height essentially changes the power in low-wavenumbers as a function of radius and height. While we acknowledge this, in this study, we stick to the choice of the surface-minimum pressure as the center of the coordinate system since we rely on averaging in radius and height and our qualitative results are not affected. For the sensitivity of center selection, see \citet{ryglicki2016deeper}.}. The radial extent of the transformed inner domain was 300 km. Figure \ref{Figure1_TCintensity.jpg} shows the comparison of HWRF forecasted intensity with best-track data (or observations from the India Meteorological Department in the case of Phailin and Lehar) for TCs Phailin (initialized on 12 hours, 9th October, 2013; Figure \ref{Figure1_TCintensity.jpg}a), Lehar (initialized on 00 hours, 26th November 20130; Figure \ref{Figure1_TCintensity.jpg}b), and Hurricane Harvey (initialized on 24th August, 2017; Figure \ref{Figure1_TCintensity.jpg}c).

While cyclic starts were used for each of the forecasts (the storm center is derived from the previous cycle initiated twelve hours prior to the initial times mentioned here), only one cycle for each storm is used in this study as shown in Figure \ref{Figure1_TCintensity.jpg}. In other words, only the HWRF forecast cycles that best captured the rapid intensity changes in the TCs of interest are presented herein. The forecast length per simulation was 90 hours for Phailin and Lehar, and 126 hours for Harvey. While HWRF does a commendable job of forecasting Phailin's RI, the rate at which the RI occurs is under-predicted when compared to the observations. The forecast cycles for Lehar's RW and Harvey's RI as well as RW agree well with their respective observations. See \citet{osuri2017prediction} and \citet{Bhalachandran2019ona} for an in-depth analysis and forecast verifications of TCs Phailin and Lehar. 

% Tang Mom
\section{Appraisal of prior dynamic-thermodynamic approaches}
Previously, researchers have explained (rapid) intensity changes in TCs using dynamic and thermodynamic approaches. The thermodynamic framework widely accepted by the TC community conceptualizes the TC as a heat engine that converts the thermal energy transferred to the air from the ocean surface to the kinetic energy (KE) of the storm \citep{emanuel1986air,emanuel1991theory}. Subsequent studies using idealized numerical experiments showed how the weakening of a TC might be explained in terms of how the energy cycle was impeded resulting in less work available to drive the surface winds \citep{tang2010midlevel,riemer2010new,riemer2011simple,tang2012sensitivity}. A common theme amongst these studies was the intrusion of low moist-entropy (or equivalent potential temperature, $\theta_{e}$) air from the environment into the TC core and the subsequent disruption of the energy cycle. \citet{Bhalachandran2019ona} built on these studies and provided an additional asymmetric component to the azimuthally-averaged thermodynamic perspectives articulated in prior studies. They showed that in a sheared vortex, a juxtaposition in the azimuthal phasing of the asymmetrically distributed downward eddy flux of $\theta_{e}$ through the top of the boundary layer, and the radial eddy flux of $\theta_{e}$ within the boundary layer was essential to establish a pathway for the external low $\theta_{e}$ to intrude the inner-core and subsequently weaken the TC. 

Alternatively, the changes in TC intensity may be explained dynamically. The dynamical framework is an extension of Newton's second law where a change in the tangential wind is derived as the response to various force terms. Furthermore, using a Reynolds averaging technique, the force terms are split into an azimuthal mean and an asymmetric term that is a deviation from the azimuthal mean \citep{haynes1987evolution,van2008tropical,persing2013asymmetric,montgomery2017recent,smith2017dynamics,leighton2018azimuthal}. 

The azimuthally-averaged tangential momentum equation is written as: 
\begin{equation}
\begin{split}
		\frac{\partial \langle v \rangle}{\partial t} = -\langle u \rangle \langle f + \zeta \rangle - \langle w \rangle \frac{\partial \langle v \rangle}{\partial z} -
		\langle u' \zeta' \rangle - \langle w' \frac{\partial v'}{\partial z} \rangle + \langle \frac{1}{\rho r} \frac{\partial p'}{\partial \theta}\rangle + Fr
        \label{tangential_momentum}
\end{split}
\end{equation}
In Eq. \ref{tangential_momentum}, the instantaneous (or time-averaged) quantities are partitioned into azimuthal mean (represented within angular brackets) and eddy terms (represented as primes). \emph{u}, \emph{v}, and \emph{w} represent the storm-relative radial, tangential, and vertical velocities respectively; $\zeta$, the vertical component of relative vorticity; $f$, the Coriolis parameter; $\rho$, the density; $p$, the pressure; $r$, the radius, and $\theta$, the azimuthal angle. The force terms that contribute to a rate of change in the azimuthally-averaged tangential velocity are the mean radial vorticity flux (first term on the right), the vertical advection of tangential momentum by the mean secondary circulation (second term on the right), the eddy radial vorticity flux (third term on the right), the eddy vertical advection of tangential momentum (fourth term on the right), the pressure perturbation term (fifth term on the right), and the frictional term (sixth term on the right). 

Our intent here is not to go over the various terms of Eq. \ref{tangential_momentum} or perform a budget analysis. Instead, our objective in this section is to highlight some of the challenges that may occur while diagnosing RI or RW using these dynamic-thermodynamic approaches. Specifically, we focus on the azimuthal-mean eddy flux term and how it manifests itself when analyzed as a function of the azimuth, and in the context of the thermodynamic fields. For example, \citet{leighton2018azimuthal} conducted a diagnostic analysis of Hurricane Edouard using ensemble simulations that produced RI as well as non-RI members in a sheared environment. They concluded that the RI members were characterized by positive eddy vorticity flux in the mid-upper regions and the non-RI members had a strong negative eddy vorticity flux in the same region. However, we find that in certain cases, an individual asymmetric signature (such as the eddy vorticity flux) is insufficient information to anticipate or forecast a forthcoming RI or RW. Such scenarios may occur when there are counteracting asymmetric variables juxtaposed with one another.     

\begin{figure}[htbp]
 \centerline{\includegraphics[width=\linewidth]{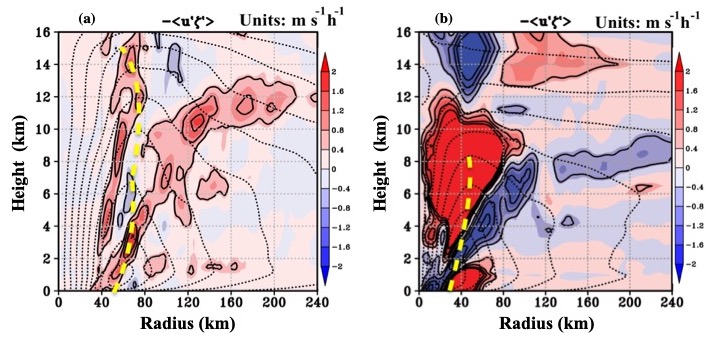}}
  \caption{Radius-height plots of eddy radial vorticity flux for Phailin (Panel a; Time-averaged during the initial period of RI - viz. 27-39 hours) and Lehar (Panel b; Time-averaged during the initial period of RW - viz. 24 - 36 hours). The red shaded regions represent a positive contribution towards intensity change and the blue shaded regions represent a negative contribution. Units for both panels are $m\;s^{-1} h^{-1}$ and the contour intervals are 0.5 $m\;s^{-1} h^{-1}$. The dotted lines represent the $\langle v \rangle$ contours and the yellow dashed line represents the radius of maximum $\langle v \rangle$ at each height.}
\label{RZeddyvorticity}
\end{figure}

Figure \ref{RZeddyvorticity} contrasts the radius-height plots of the eddy radial vorticity flux term (Third term on the right in Eq. \ref{tangential_momentum}) for TCs Phailin and Lehar. In Phailin, the eddy radial vorticity fluxes contribute positively to the spin-up of the vortex along two distinct radially outward-sloping regions that extend to heights well above the boundary layer (BL) (Figure \ref{RZeddyvorticity}a). The two distinct wall-like regions correspond to Phailin's concentric eye-walls during this period (cf. Suppl. Fig. 1). In contrast, the asymmetries in different regions within Lehar's vortex contribute positively or negatively during the same period. There is a combination of a positive eddy vorticity flux straddling the RMW in Lehar's mid-levels (between 4-12 km in the vertical), and a belt of negative eddy vorticity flux underneath it (Figure \ref{RZeddyvorticity}b). Recall that Lehar was a sheared storm that went on to experience RW. Contrary to the findings of \cite{leighton2018azimuthal}, our results show that a sheared TC can experience \textit{RW} despite an intense concentration of \textit{positive} eddy vorticity flux in the same region (cf. their Figures 8d, 9d). This suggests that mid-level positive eddy vorticity flux cannot be used as a unique signature of sheared-RI storms. Rather, our analysis seems to suggest that such a positive eddy vorticity flux is a characteristic of sheared TCs in general and that there are other counteracting mechanisms that are possibly not captured by the present framework that differentiate a sheared TC that goes on to undergo RI or RW.

\subsection*{Dynamic-thermodynamic counteractions in Lehar}
\begin{figure}[htbp]
  \centerline{\includegraphics[width=\linewidth]{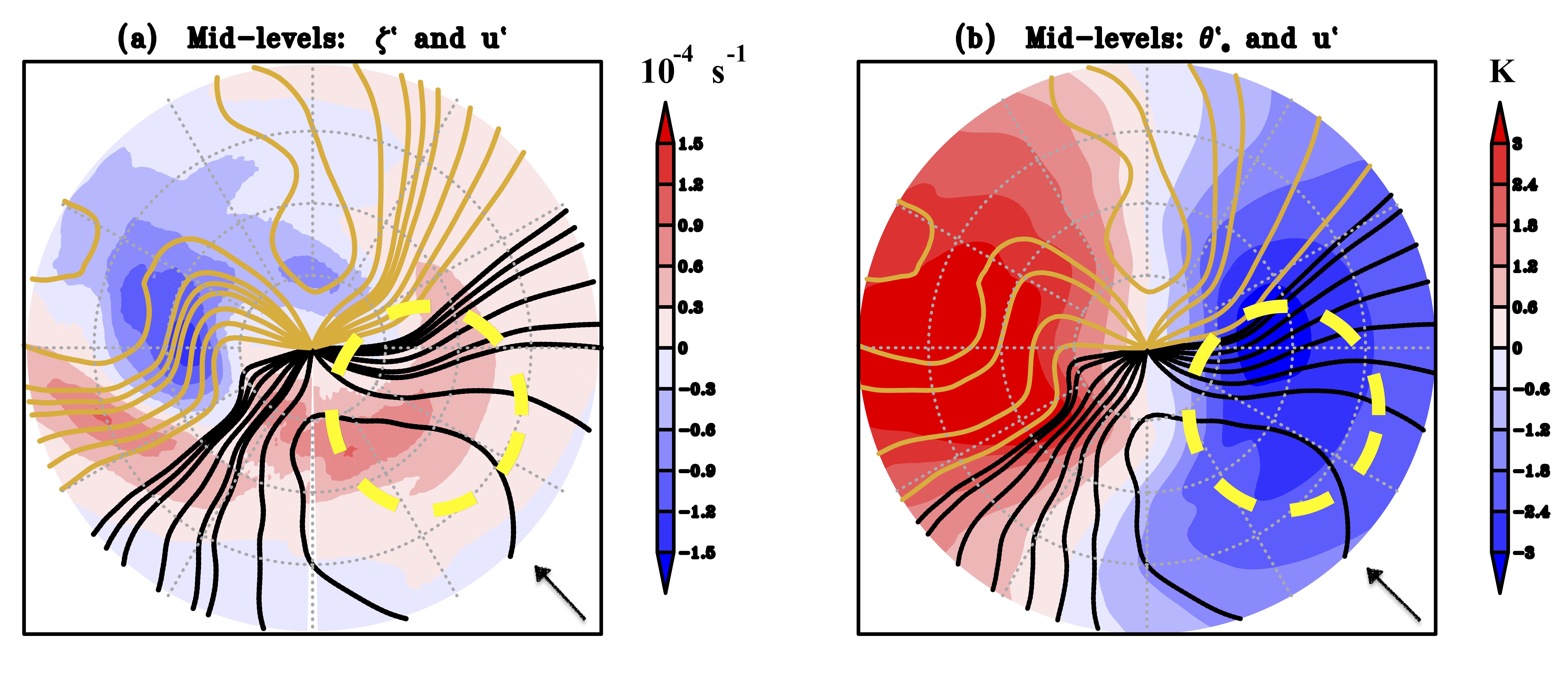}}
    \caption{Lehar's plan view (r-$\theta$) plots of (a) eddy relative vorticity (shaded) and radial velocity (contours, black represents inflow and golden represents outflow) averaged between 6-10 km (mid-levels) in the vertical, 0 - 120 km radius and 24-36 hours (b) Same as (a) except that the shading represents eddy moist entropy ($\theta_e $). Highlighted, are the regions where the inflow carries the positive eddy vorticity (a) and negative eddy $\theta_e $ (b). The arrow illustrates that the deep-shear was southeasterly during this period.}
    \label{Counter}
\end{figure}

To better understand the relationship between Lehar's eddy vorticity flux signature and its RW, we first note that the eddy vorticity flux is merely a covariance. A positive influence means a superposition between the inflow and positive vorticity or outflow and negative vorticity and vice-versa. Figure \ref{Counter} compares the horizontal cross-sections of the eddy vorticity (Figure \ref{Counter}a) with its thermodynamic counterpart - eddy moist-entropy (Figure \ref{Counter}b). Also shown are the eddy radial velocity contours. We see that for radii $\leq$ 72 km, inflow (black contours) is strongly correlated with positive eddy vorticity in the downshear region and the negative eddy vorticity is correlated with outflow (golden contours) in the upshear region (Figure \ref{Counter}a). However, for the same region, the inflow is strongly correlated with \emph{negative} eddy moist-entropy (Figure \ref{Counter}b). The eddy vorticity field is completely out of phase with the eddy moist entropy field. Under such a scenario, there is a counteraction of the dynamical process by the thermodynamical processes. This is an important finding as it reveals that the RW may occur in a sheared environment even if the conditions are dynamically favorable.  

As shear reorganizes the fields of vorticity, $\theta_{e}$, radial velocity, and the convective upward/downward motions in the azimuthal direction \citep{chen2006effects, Bhalachandran2019ona}, looking for signatures in the individual fields might be misleading given the competing nature of the mechanisms associated with these fields. Instead, we must attempt to understand the juxtaposition of these asymmetric fields and how their behavior evolves in the context of one another and the external environment. 

Moreover, the dynamic and thermodynamic impacts of shear and the intruding low $\theta_{e}$ air cannot be isolated. For example, it is unclear as to how the thermodynamic process of weakening of the heat engine (Lehar in this case) due to low $\theta_{e}$ air is reflected in the dynamics of the storm. If the dynamical perspective offered in Equation \ref{tangential_momentum} completely describes the tendency of tangential momentum, which term of Equation \ref{tangential_momentum} does the low $\theta_{e}$ air impact negatively? Such a question cannot be answered trivially without a framework that unifies the dynamic and thermodynamic effects\footnote{One possible explanation is that the ingestion of low $\theta_{e}$ air first acts to reduce the static instability within the BL and has a direct impact on the strength of the updrafts that transport the tangential momentum in the vertical direction. During the initial period of weakening, a reasonably strong (relative to the tangential wind) inflow is required to transport the low $\theta_{e}$ air into the vortex core against the tangential wind. From this perspective, we speculate that the terms to first show negative tendencies of low $\theta_{e}$ air intrusion are the mean and eddy vertical flux terms (Terms 2 and 4 in Eq. \ref{tangential_momentum}). Since the downdrafts that transport the low $\theta_{e}$ air are constrained in the azimuth to specific quadrants (e.g., upshear left in the case of Lehar, see \citealt{Bhalachandran2019ona}), it is likely that the negative tendency is first visible in the eddy vertical transport term (cf. Figure \ref{RZeddyvorticity}d - radius 40-120 km, 0-5 km in the vertical). As the radial pressure gradient reduces in tandem with the reduced deep convection, and the strength of the inflow reduces in an azimuthally averaged sense, a negative tendency is seen in the mean radial advection of vorticity (Term 1 in Eq. \ref{tangential_momentum}). The above discussion (albeit speculative) is an effort to unify the dynamic and thermodynamic perspectives.}. This motivates the need for a more precise approach that would seek to integrate the dynamic and thermodynamic effects more seamlessly to account for nonlinear, competing processes that act simultaneously.

\section{An alternative multi-scale energetics framework}
An alternative approach to the dynamic-thermodynamic frameworks presented in the previous section is an energetics approach where there is no separation between the dynamical and thermodynamical factors, and where the eddy effects may be decomposed across multiple length-scales. For example, regardless of whether the cause of Lehar's weakening was the influx of low $\theta_{e}$ or negative $\zeta$, the spin-down of a vortex may be seen as a result of the lack of sufficient available potential energy (APE) or the conversion of the same to KE. Furthermore, instead of condensing the effect of all the asymmetries into one prime term, we may classify the asymmetries based on their length-scales in the spectral domain \citep{saltzman1957equations}. Here, we compute the energetics of asymmetries at and across individual length-scales and then classify them into low-WN (WNs 1,2) or vortex-scale asymmetries, and high-WN (WNs $\geq$3) or sub-vortex-scale asymmetries. WNs 1 and 2 in addition to WN 0, explain around 85 percent of the total azimuthal variance in the APE \citep{krishnamurti2005hurricane}. In other words, WNs 0,1,2 are those that have the maximum impact directly at the vortex-scale. In the following section, we first present a power spectral analysis to compare the distribution of variance in the convective signature and KE amongst the asymmetries at different WNs at two different snapshots (that represent organized and disorganized phases respectively) from the TC's lifecycle. 

\section*{Power spectral Analysis}
\begin{figure}[htbp]
\centerline{\includegraphics[width=0.7\linewidth]{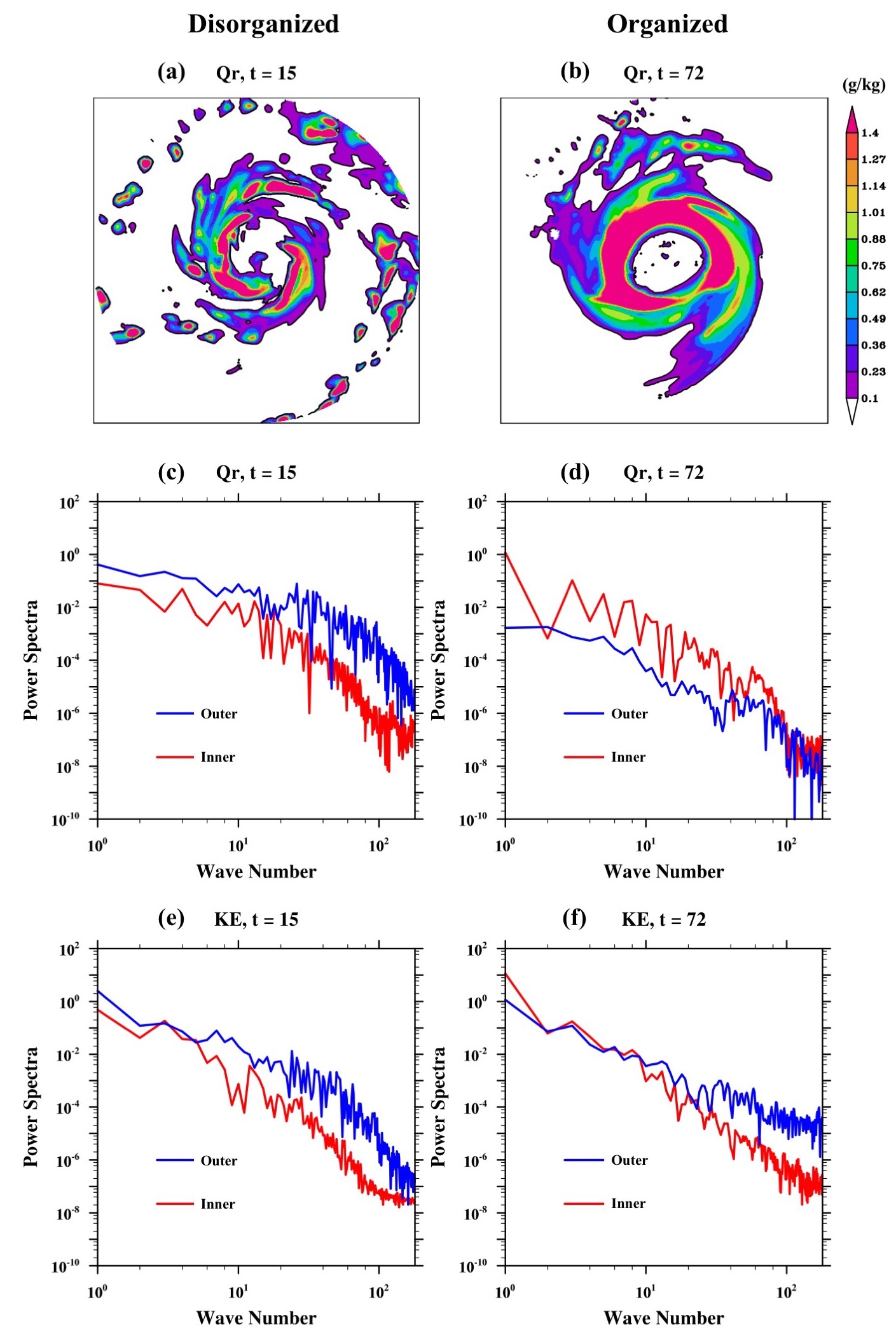}}
 \caption{(a,b) Plot of Phailin's vertically integrated cloud-water mixing ratio ($Q_r$) at t = 15 hours during a disorganized phase (a) and at t = 72 hours during its peak intensity. (c,d) present a power spectra of $Q_r$ as a function of wave number (WNs 1 to 180; The $d\theta$ here is 1 degree; As a result, there are 360 WNs out of which 180 are unique and the rest are their complex conjugates) corresponding to panels (a) and (b) averaged radially between 0-200 km (inner region) and 200-300 km (outer rain band region). (e,f) show a power spectra of kinetic energy for the times corresponding to (c,d). The axes are plotted in a logarithmic scale to show the differences in the order of magnitude across the wavenumbers. }
 \label{fig:power}
\end{figure}

Rainwater mixing ratio ($Q_r$) in high-resolution, near-cloud resolving modeling outputs, carries the \textit{aggregate signature} of individual and organized meso-convective elements and is used here as a proxy for convection\footnote{Alternatively, one may use reflectivity, vertical velocity, vertical mass flux, or diabatic heating as a proxy for aggregate convection} (Figures \ref{fig:power}a, \ref{fig:power}b). An illustration of Phailin's power spectra of the variance in $Q_r$ and eddy KE as it transitions from a disorganized phase to an organized phase is presented in Figures \ref{fig:power}c-\ref{fig:power}f.  

During Phailin's disorganized phase (at t=15, Figure \ref{fig:power}c), the power in the first ten WNs is of the same order of magnitude. As Phailin gets more organized (Figure  \ref{fig:power}d), there is an increase of power in the lower WNs and a sharp decrease in power in the higher WNs within the inner-core. A possible explanation is that as the initially disorganized vortical convective elements at outer radii aggregate, more vertical mass flux is generated. As the radial pressure gradient in the boundary layer increases in response to this removal of mass, there is convergence within the boundary layer \citep{smith2010hurricane}. The reduction of power in the outer radii (blue line) by several orders of magnitude, and the corresponding increase of power in the lower WN (particularly WN 1) is potentially an indication of such convergence (Figure \ref{fig:power}d). Additionally, as a result of aggregation and convergence, the power in WNs 10 and higher located in the outer rain band region drops significantly (roughly four orders of magnitude). 

Figures \ref{fig:power}e and \ref{fig:power}f show the power spectra in KE for the same times. Between the two periods, there is an order of magnitude increase in the KE of WN 1 and a persistent decrease in the powers of the higher WNs within the inner core region. Unlike the power spectra in $Q_r$,  there is an increase in KE in the higher WNs in the outer rain band region (Figures \ref{fig:power}e and \ref{fig:power}f). A similar analysis of the (observed) KE spectra where the slope becomes steeper as the TC intensifies can be found in \citet{vonich2018hurricane}.

While the power spectral analysis reveals the distribution of energies across scales at a given time, a more sophisticated framework is required to precisely describe the various energy transactions that add or subtract the energies at each scale and time.  

\section*{Scale Interactions}

Scale interactions is a formalism that describes the different pathways of energy exchange in the WN domain \emph{at,} and \emph{between} asymmetries of various length-scales. \citet{saltzman1957equations} laid the foundation for this framework to study global energetics and used a spherical coordinate system. Subsequently, studies such as \citet{dubey2018scale} have used the same to study the energy exchanges between synoptic-scale zonally averaged flows and associated waves (e.g., the Madden-Julian Oscillation). \citet{krishnamurti2005hurricane} retailored Saltzman's equations for a TC, by casting the system of equations in a storm-centric, cylindrical coordinate system thereby enabling the study of azimuthal asymmetries. Along the same lines, we symbolically list the different types of energy exchanges that can impact the KE of the mean flow (Equation \ref{meaneq}) and asymmetries at any scale (Equation \ref{eddyeq}). 

\begin{equation}
		\frac{\partial K_0}{\partial t} = -\sum_{n=1}^N <K_0 \rightarrow K_n> + <P_0 \rightarrow K_0> - <K_0 \rightarrow F_0>
        \label{meaneq}
\end{equation}

    \begin{equation}
    \frac{\partial K_n}{\partial t} = <K_0 \rightarrow K_n> + \sum_{ \substack{k,m \\ \hspace{-0.6cm} \pm \, k \, \mp \, m \, = \,n}} <K_{k,m} \rightarrow K_n> + <P_n \rightarrow K_n>
		- <K_n \rightarrow F_n>
        \label{eddyeq}
\end{equation}
Here, the $K_0$ and $K_n$ represent the KE of WN 0 (mean flow) and wave number \emph{n} respectively. $K_m$ and $K_k$ are the kinetic energies of WNs $m$ and $k$ that interact with WN $n$. $P_0$ and $P_n$ represent the potential energy of WN 0 and wave number \emph{n}.  The terms in angular brackets indicate exchanges that are positive in the direction of the arrow. Equation \ref{meaneq} shows that the KE of the mean flow (WN 0) could either change due to the transactions of KE between eddies of various scales, due to the conversion of APE to KE on the scale of the mean flow, and frictional dissipation  ($F_0,F_n$). The barotropic exchange of KE between mean and eddy (term 1 in Eqn. \ref{meaneq}) invokes the covariance between the mean flow and the eddy flux of momentum. Whereas, the baroclinic conversion of energy from APE to KE (term 2 in Eqn. \ref{meaneq}) invokes the covariance between vertical velocity and temperature (vertical overturning). Energy exchanges that occur at individual scales and arise due to quadratic nonlinearities are known as in-scale exchanges. 

Term 1 in Eqn. \ref{eddyeq} is the same as Term 1 in Eqn. \ref{meaneq} except that it has the opposite sign (mean's loss is eddy's gain). Term 2 in Eqn. \ref{eddyeq} represents the nonlinear exchange of KE among different scales. Energy interactions between the eddy scales occur via triad interactions that follow certain trigonometric rules. For example, WNs $n$, $m$, and $k$ can interact if and only if $k + m = n$, $-k + m = n$, or $k -m = n$. These exchanges invoke triple products and are known as cross-scale interactions. Term 3 in Eqn. \ref{eddyeq} represents the baroclinic exchanges between APE to KE at each of the WNs (\emph{n} =1 to \emph{N}) and the last term is the loss to friction. The complete equations of the above exchanges can be found in the appendix (Equations 4 to 17) and in \citet{krishnamurti2005hurricane}.

The following methodology is adopted to characterize the energetics of asymmetries during TC rapid intensity changes using the formalism of scale interactions. We take high-resolution HWRF output of the desired storms of interest and project the necessary variables on to a storm-centric, cylindrical coordinate system. We then perform a Fourier transform, and classify the resulting azimuthal harmonics of these variables into three categories: Mean (WN 0); low-WN (or large-scale) asymmetries representing the eddies that are organized at the vortex-scale and are persistent in time (WN 1,2); and higher WNs (or sub-vortex-scale) that are representative of events that are local and transient in nature (WNs 3 and higher). We then compute the energetics described in Equations \ref{meaneq} and \ref{eddyeq} during periods of rapid intensity changes. Finally, we comment on how the characteristics of multi-scale asymmetries are different during RI and RW.\footnote{As with any diagnostic analysis, what we present as results is merely the relative contribution of different processes at a few instances of time. While the analyses capture the aggregate signature and the manifestation of the processes at the period, they do not reveal the actual processes themselves.}

We formulated two initial hypotheses before performing the computations: (i) the magnitude of APE to KE conversion must be considerably higher during RI as opposed to RW, (ii) the direction of transfer must be from mean to eddy during RW and from eddy to mean during RI. The second hypothesis was formulated keeping the insights derived from large-scale general circulations in mind (e.g., \citealt{starr1970negative}). Recall that while viscous stresses always act to remove energy from the flow and transfer it to internal energy, eddies may transfer the energy back to the mean \citep{george2013lectures}. 

At this juncture, we wish to stress that our emphasis is on presenting the approach rather than the case-studies themselves. Therefore, we will use Phailin as the primary case-study in the following sections and detail each of the different energy pathways. We will use Lehar and Harvey only for contrast as and when deemed necessary. Multi-panel time-series and radius-height plots of all the terms (during periods of rapid intensity change) for all the three TCs are provided as supplementary figures.  

\subsection{Generation of APE}
\begin{figure}[htbp]
  \centerline{\includegraphics[width=\linewidth]{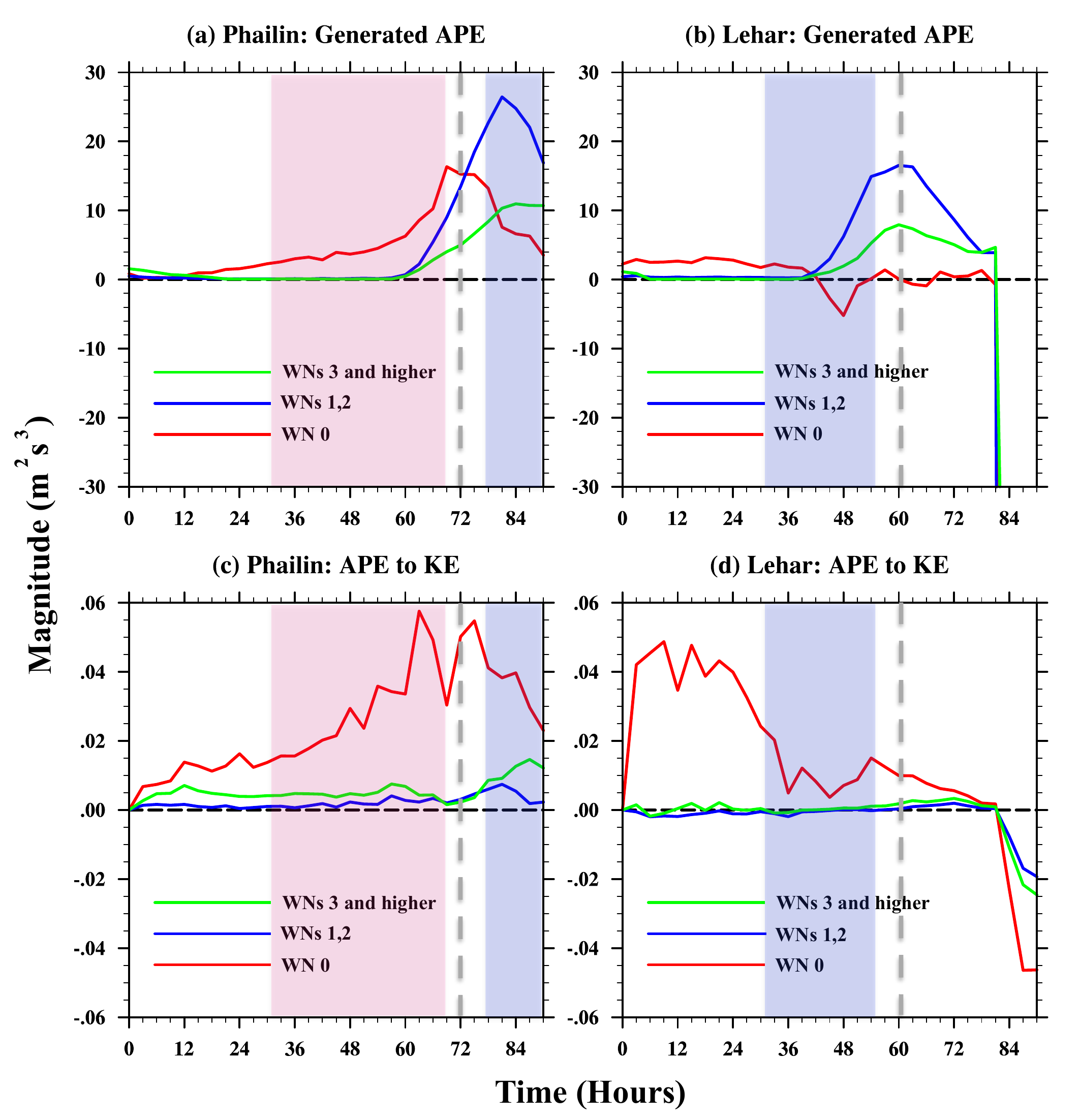}}
    \caption{Time-series of the domain-averaged (up to a radius of 300 km, up to 20 km in the vertical) rate of change of available potential energy (a,b) generated at WN 0, WN 1-2 and WNs $\geq$3. (c,d) represent the conversion from APE generated to kinetic energy. The left panels are for Phailin, and the right panels are for Lehar. The RI period is shaded in red and the RW period is shaded in blue. The landfall time is marked by a dashed, gray line.}
    \label{fig:APE}
\end{figure}

The generation of APE is a quadratic term just like the baroclinic exchanges described above. This term is computed using the covariance of diabatic heating (\emph{H}) and temperature (\emph{T}) \citep{lorenz1955available}. Such a generation can happen at every individual scale (in-scale exchange). In other words, every scale can contribute directly to the generation of APE. Figure \ref{fig:APE} shows the plot of the APE averaged within the vortex (y-axis) for each forecast time (x-axis) for the WN0, WNs 1-2, and WNs $>$ 2 for TCs Phailin and Lehar.

During Phailin's RI period, there is an increase in the mean generation of APE (red, solid line in Figure \ref{fig:APE}a), while the generation term in the asymmetries (blue and green solid lines in Figure \ref{fig:APE}a) is relatively small. However, just before landfall (as the outer rain bands of the TC vortex start interacting with land), there is an increase in the generation of APE both in the lower and higher WNs. An increase in WN 0 implies that during Phailin's RI period, the generation of APE comes from axisymmetric, organized convection of the clouds along the azimuth. However, as the storm structure is disrupted due to interaction with land, the energy is transferred to the asymmetries. In the case of Phailin, the eddies maintain the storm for a certain period during and post-landfall. Note that between the forecast time of 72 to 84 hours, the amount of APE generated in WNs 1,2 is more than twice the amount generated in the mean.   
 
An accurate treatment of TC energetics during or post-landfall will require an explicit treatment of the frictional term and this is out of the intended scope in this study. Rather, our motivation here is to show the similarities in the evolution of APE during Lehar's weakening over the ocean, and Phailin's weakening over land. Figure \ref{fig:APE}b shows the time-series plots of the generation of APE for TC Lehar. In Lehar, for the first $\sim$36 hours, the magnitude of APE generated across all WNs is comparable to that of Phailin's. However, at around 36 hours (more than 24 hours before landfall), there is a marked decrease in the mean generation of APE (red, solid line in Figure \ref{fig:APE}b) and a consequent increase in the generation of eddy APE (green and blue lines in Figure  \ref{fig:APE}b) almost 16 hours before landfall ($\sim$42 hours). Thus, regardless of whether the RW happens over the ocean or land, there is a reduction in the symmetry of APE generation. Towards the end of the RW period, the generation term across all WNs drops substantially (see e.g., 64-84 hours in Figure \ref{fig:APE}b).  

\subsection{Conversion from APE to KE}
The mechanism of transformation from APE to KE in a TC is the baroclinic overturning circulation (warm air rising and cold air sinking). This term is computed as the covariance between vertical velocity (\emph{w}) and temperature (\emph{T}). This conversion of energy happens at every individual scale through the vertical advection of high tangential momentum from the BL to the rest of the TC vortex.

However, previous studies have identified that there are preferred regions where the generated heating is more suited for conversion to KE \citep{anthes1968generation}. For example, \citet{miyamoto2015triggering} showed that when the heating associated with the organized cumulus convection occurs with the RMW, a region of high inertial stability and low static stability, there is a higher likelihood of conversion from APE to KE\footnote{\citet{smith2016efficiency} offer an alternative interpretation of heating occurring within the RMW using the evolution of angular momentum (M)-surfaces. As heating occurs within the RMW, it leads to increased convergence and spin-up within the BL; and a subsequent increase in vertical advection of high momentum within the eyewall. This results in the M-surfaces being drawn closer and spins-up the regions above the BL as well. Regardless of the explanation as to why diabatic heating within RMW is more conducive for spin-up, our emphasis is on the fact that for the same amount of APE generated, there are preferential regions within the TC vortex that are most suited to the conversion of APE to KE.}. Figures \ref{fig:APE}c and \ref{fig:APE}d show that during the RI period, there is a persistent increase in the magnitude of mean APE to KE conversion. Likewise, during the RW period (blue region) post-landfall, there is a clear reduction in the magnitude of mean APE to KE conversion due to a decrease in the ability to sustain deep convection. 

\begin{figure}[htbp] 
  \includegraphics[width=\linewidth]{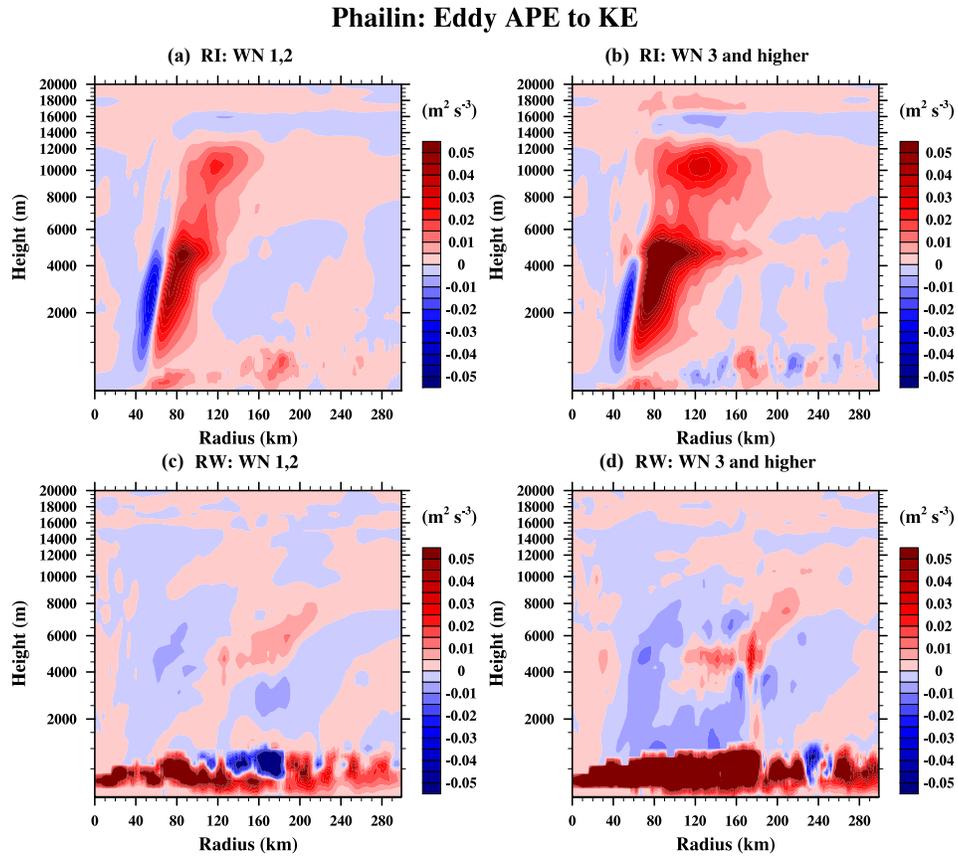}
    \caption{Phailin: Radius-height plots of the conversion from eddy potential to eddy kinetic energy for WNs 1,2 (a,c) and WNs 3 and higher (b,d). The RI and RW periods in Phailin correspond to the red shaded region and blue shaded region in Figure \ref{fig:APE}c respectively.}
    \label{fig:eddyPEKE}
\end{figure}

Since the spatially-averaged eddy terms in Figures \ref{fig:APE}c and \ref{fig:APE}d are an order of magnitude smaller than the mean terms, Figure \ref{fig:eddyPEKE} shows the radius-height plot of the magnitude of conversion from eddy APE to eddy KE during Phailin's RI (Figures \ref{fig:eddyPEKE}a, \ref{fig:eddyPEKE}c) and RW (Figures \ref{fig:eddyPEKE}b, \ref{fig:eddyPEKE}d) periods. Figure \ref{fig:eddyPEKE} reveals that during the RI period, there is a region of strong positive correlation between $w'$ and $T'$ (40-80 km radii in Figures \ref{fig:eddyPEKE}a, \ref{fig:eddyPEKE}b). This positive correlation is the aggregate signature of organized and disorganized updrafts that extend to the entire depth of the vortex within or near the eyewall region. Note that the peaks in the magnitude of the conversion from APE to KE in the eddy scales are comparable to averaged magnitudes of the conversion in the mean term in Figure \ref{fig:APE}c. Conversely, during the RW period, the ability to sustain deep convection is indicated by the positive transfer from APE to KE restricted to regions within the BL. Above the BL, there is a strong negative correlation (blue shaded region in \ref{fig:eddyPEKE}c, \ref{fig:eddyPEKE}d) between $w'$ and $T'$ indicating the vortex's inability to transfer the tangential eddy momentum from the BL to the mid and upper portions of the vortex. Therefore, as per our first hypothesis, we see that during RI, there is an increased generation of APE from heating and an increased conversion from APE to KE, and vice-versa during RW both at the mean and eddy scales. 

\subsection{Mean to eddy KE transfer}
This section focuses on the barotropic mean-eddy exchange in KE between the mean and low WNs, and the mean and high WNs. The exchange invokes the covariance between the azimuthally averaged flows and the eddy convergence of momentum (predominantly within the BL). In the KE budget formulation using the standard Reynolds averaging partitioning, this transaction between the mean and eddies would feature as a production term (e.g., see Chapter 5 of \citealt{stull2012introduction}). Since the formalism of scale-interactions gives us the flexibility to look at individual scales of asymmetries, it is important to appreciate that the transport of momentum between each eddy scale of WN $n$ and WN 0 is a unique and distinct transaction \citep{saltzman1957equations}. In this section, particular emphasis is laid on the direction of transfer between the mean and the asymmetries.

\begin{figure}[htbp]
  \centerline{\includegraphics[width=0.7\linewidth]{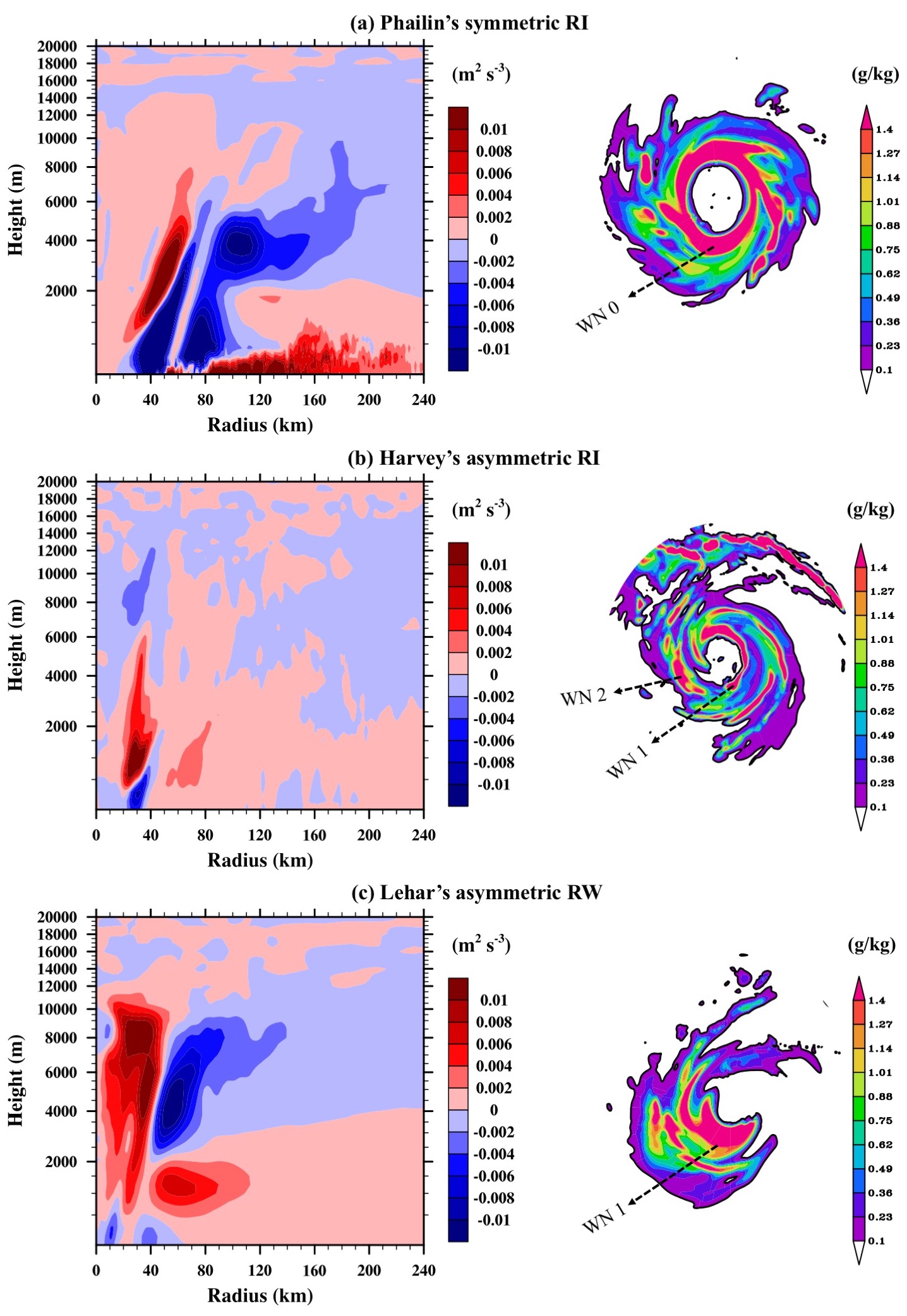}}
     \caption{(Left) Radius-height plots of the barotropic exchange between the mean and low-WN asymmetries for TC Phailin and Hurricane Harvey during their RI phases, and for TC Lehar during its RW phase. Blue shaded regions are the locations where the direction of energy transfer is from eddy to mean and red shaded regions are locations where the direction of energy transfer is from mean to eddy. (Right) Corresponding plots of rainwater mixing ratio (in g/kg) to illustrate the symmetric or asymmetric nature of convection during the highlighted periods.} 
    \label{fig:MtoL}
\end{figure}

Figure \ref{fig:MtoL} shows the radius-height plots of the KE exchange between WN 0 and WNs 1,2 and corresponding $Q_r$ plots for Phailin during its RI phase and for Lehar during its RW phase. In addition to TCs Phailin and Lehar, the same plot is shown for Harvey that underwent an asymmetric RI to contrast symmetric and asymmetric RI. The $Q_r$ plot during Phailin's RI (Figure \ref{fig:MtoL}a) indicates that Phailin's RI was characterized by azimuthally symmetric convection. During this period, between 30-80 km radii, there is a strong transfer of energy from the eddies to the mean (blue regions in Figure \ref{fig:MtoL}a). This is quite different from Harvey's RI that is supported by asymmetric convection characterized by WNs 1 and 2 (Figure \ref{fig:MtoL}b). During this period, the radius-height plot indicates that the direction of energy transfer is from the mean to the eddies (predominant regions of red in Figure \ref{fig:MtoL}b). 

The RI configurations in Figures \ref{fig:MtoL}a and \ref{fig:MtoL}b are then compared to the RW configuration in Lehar (Figure \ref{fig:MtoL}c). The $Q_r$ plot during Lehar's RW in a sheared environment indicates that the weakening period was characterized by a predominance of WN 1 convective asymmetry. During this period, there is a strong transfer of KE from the mean to WNs 1 and 2 (red regions). At larger radii within the BL, the direction of transfer is always from mean to the (organized) eddies suggesting a predominance of WN 1 and 2 asymmetries in the rain-band region. 

The dominance of mean to eddy transactions during both Harvey's RI and Lehar's RW suggests that organized asymmetries that receive energy from the mean may be associated with either RI or RW. While this transaction between the mean and low-WN eddies acts as a useful indicator as to whether the intensification or weakening process is symmetric or asymmetric, it is not very useful for the purpose of distinguishing RI from RW.

\begin{figure}[htbp]
  \centerline{\includegraphics[width=0.7\linewidth]{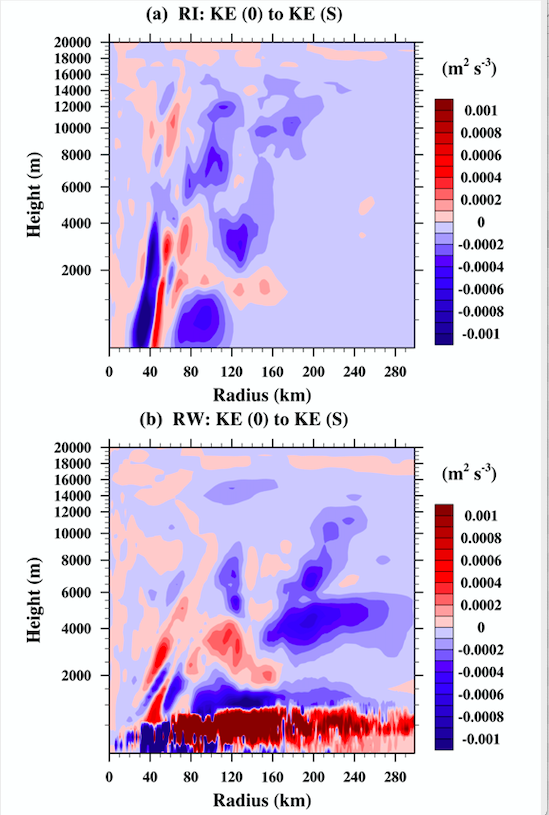}}
    \caption{Radius-height plots of the barotropic exchange between the mean and WNs $\geq$3 for TC Phailin (2013) during its RI and RW phase. }
    \label{fig:MtoS}
\end{figure}

However, we find a much more promising and consistent signature in the transaction between the mean and high-WN eddies. Figure \ref{fig:MtoS} shows the radius-height plot of TC Phailin's KE exchange between the mean and high WNs during its RI (Figure \ref{fig:MtoS}a) and RW (Figure \ref{fig:MtoS}b) phases. During RI (Figure \ref{fig:MtoS}a), there is a clear transfer of KE from eddy to mean within the BL indicating an organization of convective elements. Conversely, during RW (Figure \ref{fig:MtoS}b), there is a transfer of energy from mean to eddy within the BL indicating a disruption in the organization of convection. 

In conjunction with our second hypothesis, our findings suggest that the mean-eddy transactions during RW and eddy-mean transactions during RI are only prominent between the mean and the \textit{high-WN} eddies. It does not necessarily hold true for the low WNs. This further validates our separation of asymmetries into vortex-scale and sub-vortex-scales (or low and high WNs) suggesting that they may behave differently.

The asymmetries at the smallest scales play an additional major role in dissipating the turbulent KE into internal energy \citep{kolmogorov1941degeneration}. The direction of transfer between mean and eddy in Figure \ref{fig:MtoS} is simply the bulk signature of the organization and disruption of convection that dominates the trickle-down effect (from the mean to the smallest of eddies until viscosity dissipates the energy) that occurs at all times in turbulent flows.

\subsection{Eddy to eddy KE transfer}
This section addresses the cross-scale interactions between the low-WNs and high-WNs. These exchanges are computed using triple products since the exchanges take place as triads (WN $n$ with WNs $m$ and $k$). Importantly, these transactions indicate how the smaller scales project onto the larger-scale through upscale or downscale transfers. 

\begin{figure}[htbp]
  \centerline{\includegraphics[width=\linewidth]{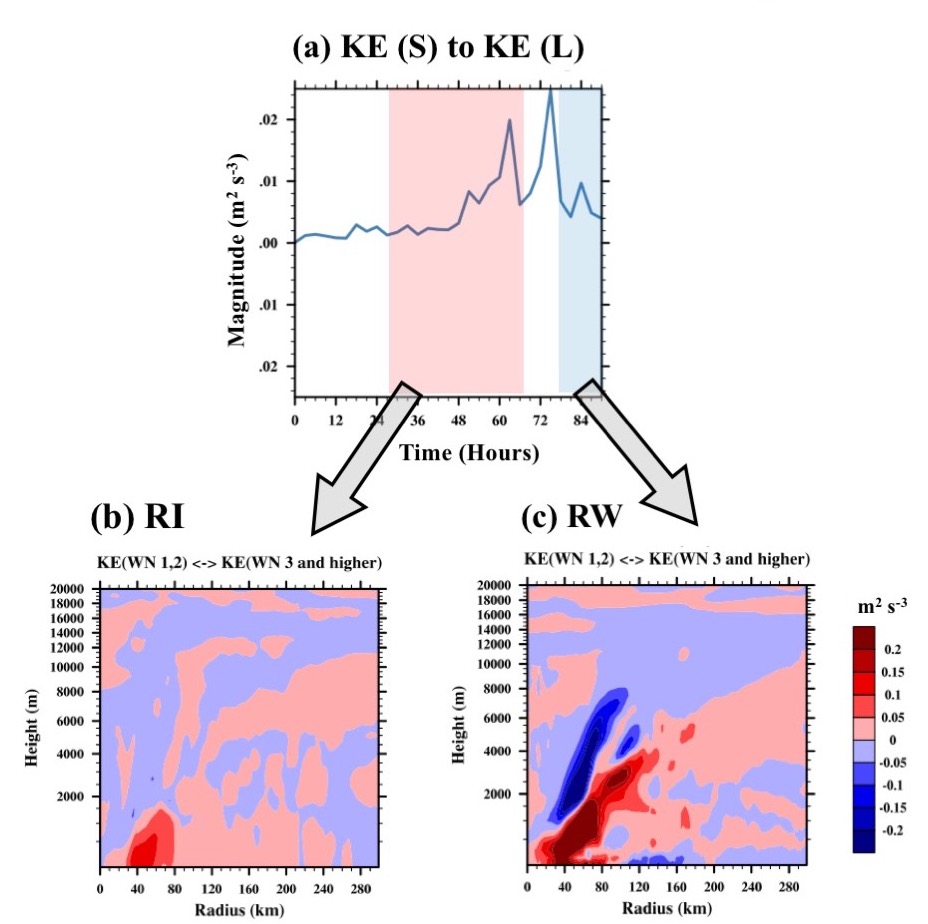}}
    \caption{(a) Time-series plot of Phailin's (domain-averaged) rate of change of kinetic energy of low-wavenumbers due to eddy-eddy interactions with the high-wavenumbers. In this figure, an increase represents an upscale transfer of KE from small to large scales, and a decrease represents a downscale transfer of KE from large to small scales. (b) and (c) Radius-height plots of the same quantity time-averaged during the Phailin's RI period (highlighted in red in (a)) and RW phase (highlighted in blue in (a)) respectively. }
    \label{fig:triads}
\end{figure}

To understand the eddy-eddy energetics during Phailin's RI and RW periods, we present the time-series of Phailin's KE transfer from the high to low WNs (Figure \ref{fig:triads}a). During Phailin's RI period, there is an upscale transfer from the higher to lower WNs. On the other hand, during the RW period, there is a downscale transfer from the lower to higher WNs. Figures \ref{fig:triads}b and \ref{fig:triads}c are the radius-height plots of the same exchange time-averaged during RI and RW respectively. Figure \ref{fig:triads}b suggests that the majority of the upscale KE transfer occurs within the BL with the maximum values straddling the RMW. Likewise, during the weakening phase, the majority of the downscale transfer happens within the BL (radii $\geq$ 100 km), and in the eyewall region within the RMW where the diabatic heating is concentrated (Figure \ref{fig:triads}c). The upscale and downscale cascade of turbulent KE described here has also been noted previously in observational analyses of the BL \citep{byrne2013height}.

\subsection{On the near-independence of mean and eddy terms}
\begin{figure}[htbp]
  \centerline{\includegraphics[width=\linewidth]{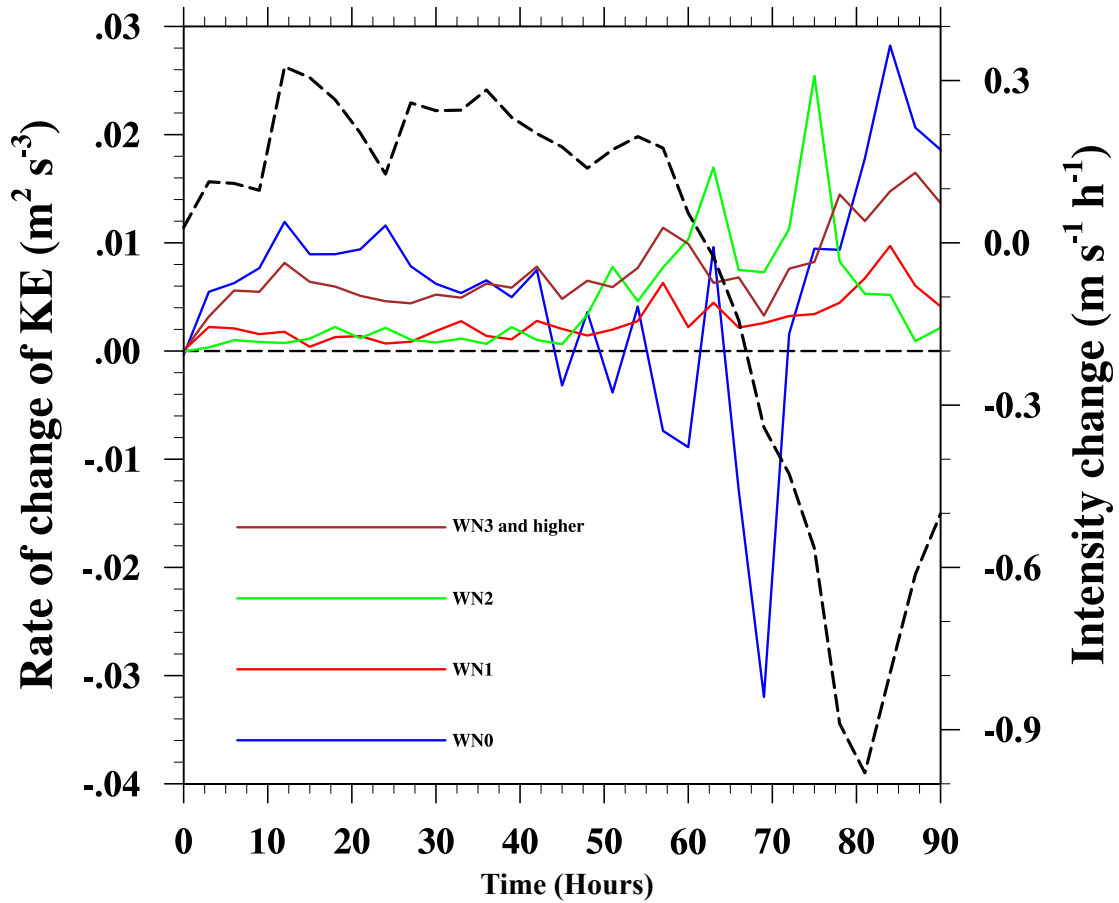}}
    \caption{Time-series of the \emph{net} change of KE in WNs 0,1,2,$\geq$3 (blue, red, green and brown solid lines) in Phailin compared against the rate of change in intensity (black dashed line) over the course of its life-cycle. This figure serves to illustrate that due to the existence of multiple modes of energy exchanges that influence the KE at a particular scale, it is entirely possible that the energies at the mean and the eddies grow at the same time (t= 10 to 30 hours) or that the eddies grow at the expense of the mean (t = 65 to 75 hours).  }
    \label{fig:meaneddy}
\end{figure}

The above discussions on the multiple competing modes of energy exchanges that influence the KE at a particular scale reveal an important message regarding the transfer of energy between the mean and eddies: the growth and decay of KE in the mean or eddy scales need not happen at the expense of one another. Figure \ref{fig:meaneddy} further illustrates this point with the time-series plot of the net rate of change in KE (computed using Equations \ref{meaneq} and \ref{eddyeq}) at WNs 0, 1, 2, and $\geq$3 over the course of Phailin's life-cycle. The change in KE in each of the scales is plotted against the rate of change in intensity (cf. with Figure \ref{Figure1_TCintensity.jpg} for the time-series of intensity) to indicate which scales had the direct impact on intensity change at a particular period in time. Between 0-36 hours, the change in intensity is correlated with an increase in KE across all the WNs (mean and eddy scales). As expected, during this period, WN 0 is the most dominant scale. However, the net KE in all the small scales is comparable to that of WN 0. At about 42 hours, the KE of the mean declines. During this period, the eddy scales maintain the TC until the intensity starts to decrease rapidly after 72 hours. Between 64-72 hours, the KE in the mean reduces significantly while the energies in the eddy scales continue to grow. Note that this is a comparison between aggregated quantities (the energetics are domain-averaged in radius and height throughout the vortex) and intensity, which is a surface-based quantity. Therefore, it is important to keep in mind that time-scale of the response of the vortex as a whole is larger than that of the surface winds.

The above discussion disproves any notion that the eddies only grow at the \emph{expense} of the mean. With this, we reemphasize that if we overlook the multiple competing pathways of energy transfer at a particular scale, the consequent results will lead us to draw incorrect conclusions regarding the very nature of eddies during TC rapid intensity changes.

\subsection{Order of magnitude analysis}
\begin{figure}[htbp]
  \centerline{\includegraphics[width=\linewidth]{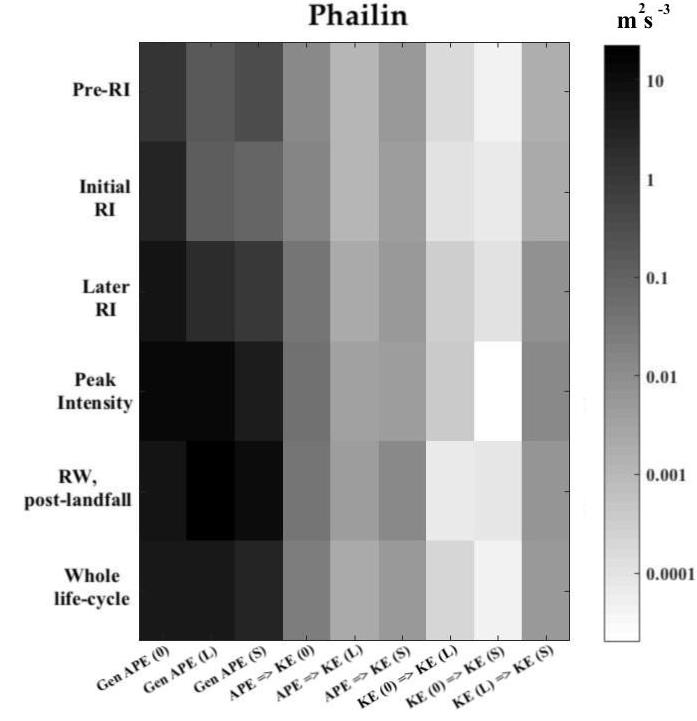}}
    \caption{Order of magnitude analysis for the various energy transactions over the course of TC Phailin's life-cycle.  }
    \label{orderofmag}
\end{figure}

Given the competing nature of the various terms that influence the energetics at and across the various scales within a TC, it is important to understand the relative importance of these terms at various stages in a TC's life-cycle. Figure \ref{orderofmag} presents such an analysis and underscores the order of magnitudes (only absolute values are presented here) of the various (averaged in radius and height within the vortex) energy transactions that take place over the course of the life-cycles of TC Phailin. The objective of Figure \ref{orderofmag} is not to highlight the differences between the different phases in a TC's life-cycle. Rather, we show that regardless of the phase of the storm’s life-cycle, the relative importance of the energetics remain fairly constant.

First and foremost, the comparison of the generation of APE in the mean (0), large-scale asymmetries (L), and small-scale asymmetries (S) indicates that the generation in each of these scales is of the same order of magnitude ($10^{1}$). As one would expect, the baroclinic conversion terms from APE to KE are one order smaller than the generation term, indicating that not all of the PE generated is converted to KE. The baroclinic conversion in the mean is one order higher than the asymmetric conversion from PE to KE ($10^{-1}$ in the mean and $10^{-2}$ in the eddy scales). Interestingly, the barotropic transactions between the mean and eddy scales are the smallest amongst all the transactions (of the order of $10^{-4}$ or $10^{-5}$). This finding may be counter-intuitive but justifiable given that a TC vortex is a largely baroclinic system due to the abundance of buoyant updrafts and corresponding downdrafts. This finding also explains why the trends in the mean and eddy terms were found to be near-independent in the previous section. 

However, the fact that the barotropic exchanges are so small compared to the baroclinic transactions raises concerns about the previous treatment of the processes of axisymmetrization using purely barotropic diagnostic techniques (e.g., \citealt{smith1995vortex,moller1999vortex,hendricks2004role,kwon2005dynamic}). Axisymmetrization or the growth of WN 0 can occur either due to symmetric, baroclinic conversion from APE to KE or due to the individual interaction of WN 0 with eddies of any WN $n$ (Equation \ref{meaneq}). Our results suggest that the process of axisymmetrization occurs when the mean (WN 0) gains a significant amount of KE directly from APE as an in-scale baroclinic transaction rather than a barotropic transaction from the eddy terms. Note that the above discussion is applicable at a system (vortex)-scale. It is possible that barotropic dynamics are important at a local scale and such an impact does not translate into a system-scale transition. 

On the other hand, the eddy-eddy (across-scale) transactions (last column in Figure \ref{orderofmag}) are of the same order of magnitude as the baroclinic terms ( $10^{-1}$ during peak intensity and RI periods and $10^{-2}$ during other periods). These eddy-eddy transactions are of crucial importance since they represent the mechanism through which localized asymmetries can feed energy into organized vortex-scale asymmetries (e.g., through the successive merging of small-scale vortices described in \citealt{montgomery2006vortical}). It must also be noted that the contributions from the higher WNs are non-negligible and should not be ignored. Given their shorter time-scales, the eddy-eddy cross-scale interactions are the primary energy pathway by which the small-scale eddies impact vortex-scale transitions. Studies using \emph{linear} models that only permit transactions between the mean and eddies (e.g., \citealt{nolan2007tropical}) or those diagnostic frameworks that follow a Reynolds averaging-based treatment of the eddy terms (e.g., \citealt{miyamoto2015triggering}) fail to capture this important mode of energy exchange through which the asymmetries influence the system-scale KE and intensity. Such treatments may result in the misinterpretation of the role of barotropic mean-eddy dynamics in the context of TC intensity changes. 

Thus, Figure \ref{orderofmag} illustrates that regardless of the phase of the storm’s life-cycle, the relative importance of the energetics remain fairly constant. However, this certainly does not imply that there is no value in analyzing the difference between the phases. Rather, the key message here is that the order of magnitude of the difference between the three energy pathways is much greater than the difference between the individual terms at various phases (RI and RW for example in Figures 5-9).    

\section{Summary and Conclusions}
This work seeks to better understand and characterize the nature of asymmetries within a TC vortex during periods of \textit{rapid} intensity changes. Prior attempts at investigating the asymmetric impacts employ a mean-eddy partitioning that condenses the effect of all the asymmetries into one term. Such techniques fail to highlight the differences in the role of asymmetries at different scales. With this in mind, we present an energetics-based approach that allows for the analysis of the impact of asymmetries at multiple length-scales in the spectral domain. Furthermore, we identify the limitations of prior approaches that separate the dynamic and thermodynamic asymmetric effects. We show an example wherein competing dynamic and thermodynamic effects make it difficult to diagnose the cause of the outcome.

\begin{figure*}[htbp]
  \centerline{\includegraphics[width=\linewidth]{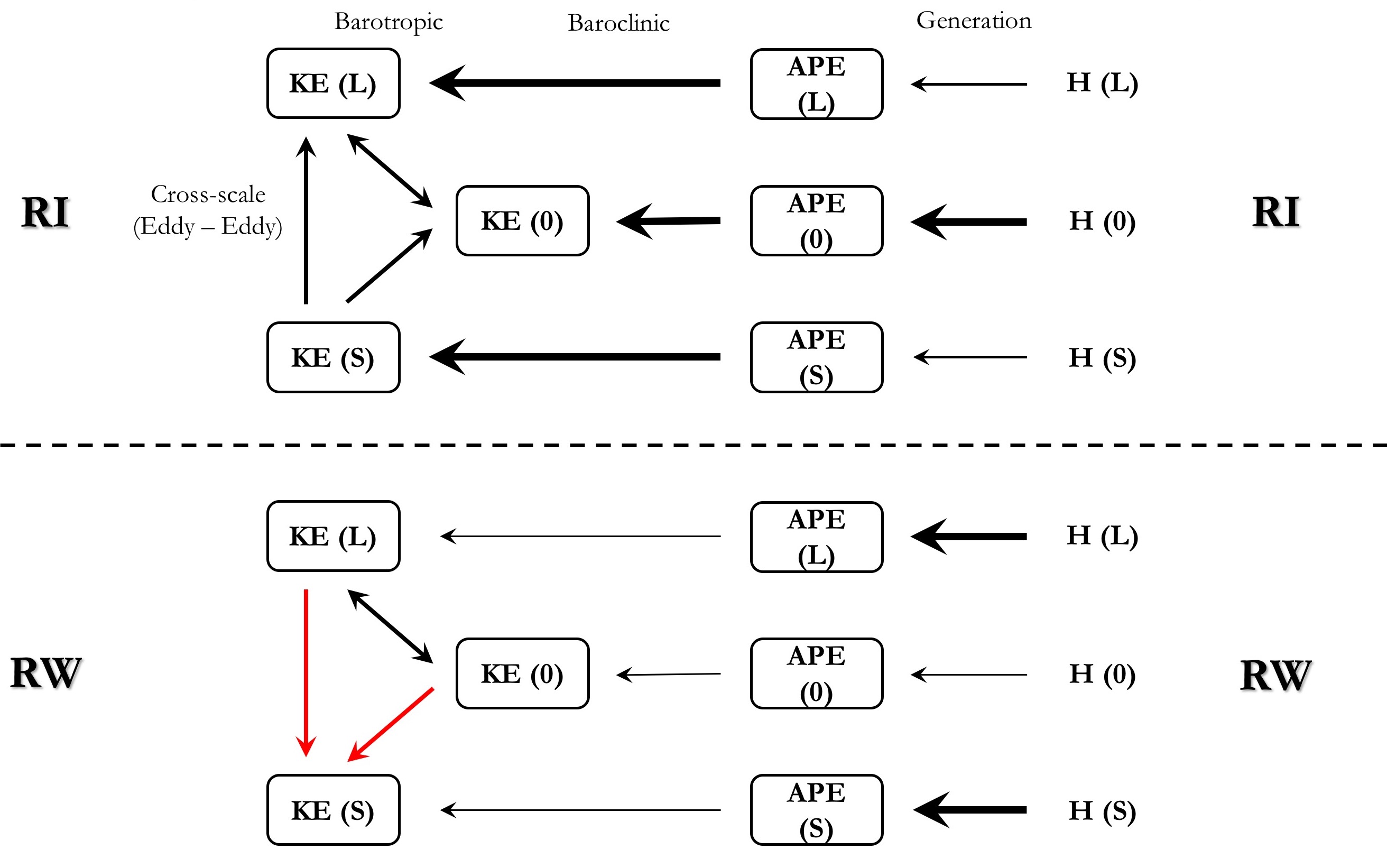}}
    \caption{Summary of the insights from scale-interactions during RI and RW phases. In this figure, the weight of the arrow corresponds to the magnitude of the energy exchange and the change in color and direction represents a change in the direction of transfer of energy. Double-sided arrows indicate that the energy transaction can be either way during RI and RW.}
    \label{fig:summary}
\end{figure*}

We then conduct a numerical investigation of TCs Phailin, Lehar, and Harvey, and identify the commonalities in the energetics of their multi-scale asymmetries during periods of rapid intensity changes. Our salient findings based on the case studies presented here are: (see Figure \ref{fig:summary} for an illustration of the same): 

\setlist{nolistsep}
\begin{itemize}[noitemsep]
    \item The APE generation term is largely symmetric (i.e., the order of magnitude of APE generation at WN 0 is the highest) during the RI period until the vortex encounters an external instability such as shear or landfall. In our case studies, we find that regardless of whether the RW happens over the ocean or land, there is a reduction in the symmetry of APE generation (decrease of APE in WN 0 and a corresponding increase in WNs 1 and 2). While we note the similarities in the behavior of the APE generation term between the RW over the ocean and landfall, we have not explicitly resolved the frictional effects in this work. We duly note that the difference in friction over land as compared to the ocean (besides that it is enhanced) is that it may become highly asymmetric and thus remove energy differently at different scales. Scale-dependent energetics inclusive of frictional dissipation is certainly something that warrants further research.
    \item There is an increased conversion from APE to KE during RI, and vice-versa during RW at all scales.
    \item A consistent signature of KE transfer from eddy to mean (mean to eddy) during RI (RW) was notable only between the mean and higher-WN (sub-vortex-scale) eddies. On the other hand, the direction of transfer between mean and the vortex-scale (low-WN) eddies was in either direction during RI and RW for different cases. This is because TCs can undergo either a symmetric RI (WN 0 is the dominant scale), where the energy transfer is from low-WN eddies to WN 0 or an asymmetric RI (WNs 1 and/or 2 are the dominant scales), where the energy transfer is from WN 0 to the low-WN eddies. Such a finding validates the separation of asymmetries into different length-scales in addition to WN 0 as opposed to a more simplistic mean and (single) eddy term.
    \item The cascade of KE is predominantly upscale (high to low WNs) during RI and downscale (low to high WNs) during RW. These cross-scale transactions of KE are important in that they tell us how the smaller, transient, and less-predictable sub-vortex scales project onto the vortex-scale eddies that are persistent in time and by extension, more predictable.
    \item The energetics of asymmetries are largely independent of the mean.
\end{itemize}
Our order of magnitude analysis shows the baroclinic and cross-scale energy exchanges are at least two orders of magnitude greater than the barotropic transactions throughout the life-cycle of the cases studied here. This result suggests that, contrary to conventional wisdom, the primary mechanism of axisymmetrization is baroclinic conversions from PE to KE directly at WN 0; and the primary mechanism of convective (dis)aggregation is the cross-scale exchanges of KE amongst the eddies of different length-scales; not the barotropic transactions that involve direct mean-eddy transactions of eddies at different length scales. Linear models and purely barotropic diagnostics may lead to incorrect conclusions regarding the importance of barotropic transactions due to aliasing. Importantly, the order of magnitude analysis helps us identify the direction of cross-scale interactions, and the magnitude of APE to KE conversion, as potential early-warning indicators of rapid intensity changes in asymmetric TC vortices.   

\subsection*{Future Work}
A logical next step is to apply the methods developed and demonstrated here to a larger set of cases to test for systematic dependencies of intensity changes on specific combinations of energy transfers. The approach and analyses delineated here represent merely the first steps towards an improved understanding of the behavior of multi-scale asymmetries and their consequent impact on TC intensity changes. In the future, we intend to extend the diagnostics presented here to multiple case-studies and perform a statistical analysis such as the linear discriminant analysis to examine the relative importance of the various energy pathways \citep{Bhalachandran2018The}. 

Furthermore, we can use the techniques described here to evaluate the representation of these energy exchanges in TC forecast models in the context of RI and RW. Such information can potentially aid in the improvement of the components in the model that directly influence the aggregation and disruption of the organization in convection (e.g., grid-scale and subgrid-scale cumulus convection, and model diffusion). This warrants for an improved understanding of the spatio-temporal aspects of the energetics. Additionally, our approach offers an important pathway to better understand the predictability and stochasticity of multi-scale asymmetries and quantify model uncertainty using ensemble model runs (See e.g., \citealt{bhalachandranconceptual}). Finally, scale-interactions is a formalism that may be extended to understanding environment-vortex interactions. In its present state, the framework can only indirectly quantify such environment-vortex interactions since our analysis is conducted purely from the viewpoint of the TC vortex. Since our present definition of ‘scales’ is hinged on a storm-centric, cylindrical coordinate system with a defined center of circulation, such a definition may not extend to the exchange of energy in the large-scale environment. To explicitly examine the scale-interactions in the environment or those between the environment and the vortex, two separate analyses will need to be conducted. One, at a large-scale focusing on zonal/meridional asymmetries (e.g., \citealt{saltzman1957equations}) and one focusing on azimuthal asymmetries. These interesting topics are out of the intended scope of the present study but are certainly worthy of continued research.

%%%%%%%%%%%%%%%%%%%%%%%%%%%%%%%%%%%%%%%%%%%%%%%%%%%%%%%%%%%%%%%%%%%%%
% ACKNOWLEDGMENTS
%%%%%%%%%%%%%%%%%%%%%%%%%%%%%%%%%%%%%%%%%%%%%%%%%%%%%%%%%%%%%%%%%%%%%
\section*{Acknowledgments} 

The authors particularly thank Dr. Ghassan Alaka and Russell St. Fleur from NOAA/HRD for providing the HWRF outputs for Hurricane Harvey. S.B gratefully acknowledges the financial support from NASA in the form of a NASA Earth Science Fellowship (Grant no: NNX15AM72H) and the Bilsland Dissertation Fellowship. This paper is dedicated to the memory of our coauthor, colleague, and mentor Prof. T. N. Krishnamurti, who died on 7th February, 2018.

\newpage 

%%% Appendix 
\section*{Appendix: Pertinent Equations for Scale Interactions}
In a storm-centric cylindrical coordinate framework, the horizontal momentum equations are given by: 

\begin{equation}
\frac{\partial v_\theta}{\partial t} = - v_\theta \frac{\partial v_\theta}{r \partial \theta} - v_r \frac{\partial v_\theta}{\partial r} - \omega \frac{\partial v_\theta}{\partial p} - \frac{v_r v_\theta}{r} - fv_r - g \frac{\partial z}{r \partial \theta} - F_\theta,
\end{equation}

\begin{equation}
\frac{\partial v_r}{\partial t} = - v_\theta \frac{\partial v_r}{r \partial \theta} - v_r \frac{\partial v_r}{\partial r} - \omega \frac{\partial v_r}{\partial p} - \frac{v^2_\theta}{r} + fv_\theta - g \frac{\partial z}{\partial r} - F_r,
\end{equation}

and the continuity equation is: 
\begin{equation}
\frac{\partial v_\theta}{r \partial \theta} + \frac{\partial v_r}{ \partial r} + \frac{v_r}{r} + \frac{\partial w}{\partial p} = 0
\end{equation}

The independent variables here are the azimuthal angle $\theta$, the radius from the storm center $r$, and the pressure $p$. The tangential, radial, and vertical components of wind are $v_\theta$, $v_r$, and $\omega$. Additionally, $f$ is the Coriolis parameter; $F_\theta$ and $F_r$ are the tangential and radial components of the frictional force per unit mass. 

The Fourier transform of any variable $ q(\theta,r,z)$ along the azimuthal direction is given by: 
\begin{equation}
q(\theta,r,z) = \sum^\infty_{n=-\infty} Q(n,r,z) e^{in\theta}
\end{equation}
where the complex Fourier coefficients $Q(n,r,z)$ for any WN $n$ are given by
\begin{equation}
Q(n,r,z) = \frac{1}{2\pi} \int^{2\pi}_{0} q(\theta,r,z) e^{in\theta} d\theta 
\end{equation}
We will follow the same notation (capital letters for the Fourier coefficients of fields represented by small letters) henceforth. The Fourier transform of the product of two functions $q_1$ and $q_2$ is given by their convolution in the WN domain: 

\begin{equation}
\frac{1}{2\pi} \int^{2\pi}_{0} q_1(\theta,r,z) q_2(\theta,r,z) e^{in\theta} d\theta = \sum^\infty_{m=-\infty} Q_1(n,r,z) Q_2(n-m,r,z)
\end{equation}

Using equation 9, equations 4 to 6 can be shown to lead to the following equation for the rate of change of kinetic energy of a given WN $n$ due to across-scale nonlinear interactions with all the other WNs other than $0$:

\begin{equation}
\begin{split}
\frac{\partial K(n)}{\partial t} = \sum_{\substack{m=-\infty \\ m\neq 0}}^\infty v_\theta(m) (\frac{1}{r} \Psi_{v_\theta \frac{\partial v_\theta}{\partial \theta}}(m,n) + \Psi_{v_r \frac{\partial v_\theta}{\partial r}}(m,n) +  \Psi_{\omega \frac{\partial v_\theta}{\partial p}}(m,n) + \frac{1}{r}\Psi_{v_\theta v_r}(m,n))  \\
  + v_r(m) (\frac{1}{r} \Psi_{v_\theta \frac{\partial v_r}{\partial \theta}}(m,n) + \Psi_{v_r \frac{\partial v_r}{\partial r}}(m,n) +  \Psi_{\omega \frac{\partial v_r}{\partial p}}(m,n) - \frac{1}{r}\Psi_{v_\theta v_\theta}(m,n)) \\
  -\frac{1}{r} \frac{\partial}{\partial r} r [ v_\theta(m) \Psi_{v_r v_\theta}(m,n) + v_r(m) \Psi_{v_r v_r}(m,n)] \\
  -\frac{\partial}{\partial p} r [ v_\theta(m) \Psi_{\omega v_\theta}(m,n) + v_r(m) \Psi_{\omega v_r}(m,n)] 
\end{split}
\end{equation}
where for fields a,b we define $\Psi_{ab}(m,n)$ = $A(n-m)B(-n) + A(-n-m)B(n)$ using A and B as their Fourier coefficients. 

For azimuthal WN 0, the generation of available potential energy is calculated as the covariances of heating ($H$) and temperature ($T$) and is expressed by 

\begin{equation}
Gen(APE(0)) = \langle H,T \rangle =  \int_V \gamma (H - \overline{\overline{H}}) (T - \overline{\overline{T}}) \, \rho dV,
\end{equation}
\begin{equation}
\gamma = - \frac{\frac{T_{pot}}{T} (\frac{R}{C_p p})}{\frac{\partial T_{pot}}{\partial p}}
\end{equation}
where $\gamma$ is the static stability parameter, $T_{pot}$ is the potential temperature, $R$ is the universal gas constant, and $C_p$ is the specific heat at constant pressure. The double overbars indicate a horizontal areal average of heating and temperature. $V$ represents the volume of integration. 

The covariance $\langle H,T \rangle$ can be broken down into its azimuthal harmonic components $0 ... n$, 
\begin{equation}
\langle H,T \rangle = \langle H_0,T_0 \rangle + \langle H_1,T_1 \rangle + ... + \langle H_n,T_n \rangle
\end{equation}
where the angle brackets represent an azimuthal mean. 

The generation of APE at any WN is then expressed by 
\begin{equation}
Gen(APE(n)) =   \int_V \gamma H_{n} T_n \, \rho dV
\end{equation}

The baroclinic, in-scale conversion from available potential to kinetic energy is computed using the covariance between vertical motion and temperature, and is given by: 
\begin{equation}
<APE(0) \, \rightarrow \, KE(0)> \, = \langle w(0),T(0) \rangle = - \int_V C_p \frac{(w - \overline{\overline{w}}) (T - \overline{\overline{T}})}{p} \, \rho dV,
\end{equation}
For all other scales, 
\begin{equation}
<APE(n) \, \rightarrow \, KE(n)> = \langle w(n),T(n) \rangle \, =   - \int_V C_p \frac{(w_n) (T_n)}{p} \, \rho dV
\end{equation}

The barotropic transactions of kinetic energy between the mean and a given eddy of WN $n$ is given by
\begin{equation}
\begin{split}
<KE(0) \, \rightarrow \, KE(n)> \, = \sum^{\infty}_{n=1} [\Phi_{v_\theta v_r} (n) \frac{\partial \langle v_\theta \rangle}{\partial r} + \Phi_{v_r v_r} (n) \frac{\partial \langle v_r \rangle}{\partial r} + \Phi_{v_\theta w} (n) \frac{\partial \langle v_r \rangle}{\partial p} \\
+ \frac{1}{r}\Phi_{v_\theta v_\theta} (n) \langle v_r \rangle - \frac{1}{r}\Phi_{v_\theta v_r} (n) \langle v_\theta \rangle]
\end{split}
\end{equation}
where for fields a,b we define  $\Phi_{ab}(n)$ = $A(n)B(-n) + A(-n)B(n)$ using A and B as their Fourier coefficients.   

%%%%%%%%%%%%%%%%%%%%%%%%%%%%%%%%%%%%%%%%%%%%%%%%%%%%%%%%%%%%%%%%%%%%%%
\pagebreak 
% \bibliographystyle{ametsoc2014}
% \bibliography{references}

%%%%%%%%%%%%%%%%%%%%%%%%%%%%%%%%%%%%%%%%%%%%%%%%%%%%%%%%%%%%%%%%%%%%%
% FIGURES---PLACE AT END OF DOCUMENT
%%%%%%%%%%%%%%%%%%%%%%%%%%%%%%%%%%%%%%%%%%%%%%%%%%%%%%%%%%%%%%%%%%%%%

\newpage
\section*{Figures}

\end{document}